\documentstyle[aps,prl]{revtex}
\title{Direct emission of light mesons from quark-gluon plasma surface.}
\author{D.Yu. Peressounko, Yu.E. Pokrovsky}
\address{Russian Research Center "Kurchatov Institute", 123182, Moscow.}
\begin{document}
\maketitle

\begin{abstract}
On the basis of hydrodynamic model of evolution of heavy ion collision we
consider emission of $\pi,K,\eta,\rho,\omega,K^*$ mesons directly from the
surface of quark-gluon plasma, with accounting of their absorption by
surrounding plasma hadronic gas. We evaluate upper and lower limits on yields
of these direct mesons in Pb+Pb collision at $158\,A\cdot GeV$ (SpS) and
$3150+3150\,A\cdot GeV$ (LHC), and find, that even in the case of the lowest
yield, direct $\eta$ and heavier mesons dominate at LHC energy at soft $p_t$
($p_t\le 0.5\,GeV/c$). This enhancement of the low $p_t$ meson production
can be considered as a quark-gluon plasma signature.
\end{abstract}

\pacs1{25.75.-q,12.38.Mh,25.75.Dw,24.10.Nz}

One of the goals of the ultrarelativistic heavy ion physics is a creation and
registration of a new state of matter -- quark-gluon plasma (QGP). However,
being created in the heavy ion collision, QGP expands, cools and transforms
back to hadronic gas, and up to now there is no reliable signal, which
identifies uniquely its creation.

As a rule, considering heavy ion collision with QGP creation, one assumes,
that hadrons, produced on plasma surface, suffer numerous rescatterings in
the surrounding hadronic gas, and thus final hadrons do not carry direct
information about plasma. In this paper we show that this widely accepted
opinion is not quite right for the case of finite systems, created in the
heavy ion collisions. Because of the comparability of free path lengths of
hadrons in the hadronic gas with sizes of region, occupied by hot matter,
significant number of hadrons, created on surface of quark-gluon plasma can
pass through the surrounding hadronic gas without rescattering. Such kind of
final state hadrons we call {\it direct} hadrons below. In contrast to
usually considered {\it freeze-out} hadrons, originated due to evaporation
from the surface of hadronic gas or its freeze-out, direct hadrons carry
immediate information about the plasma surface.

Emission of hadrons from the depth of hadronic gas was considered in
literature several times. In the frame of quark-gluon string model it was
shown \cite{Bravina}, that final hadrons, created in the heavy ion collision,
originate not from thin freeze-out hypersurface, but from the whole volume,
occupied by hot matter. Somewhat similar to our approach was used in papers
\cite{Grassi} for evaluation of hadronic emission from the depth of hadronic
gas. It was shown, that this emission is not small enough to neglect, and can
be one of the sources of enhancement of strangeness production in
ultrarelativistic heavy ion collisions. Nevertheless, to our knowledge, until
our previous paper \cite{We}, there was no special consideration of hadronic
emission directly from the surface of QGP. We would like to stress, that in
contrast to hadronic emission from the depth of hadronic gas, emission of
direct hadrons from the surface of QGP gives unique opportunity to explore
QGP properties by means of hadrons.

In our previous paper \cite{We} we considered S+Au collision at $200\,A\cdot
GeV$ (SpS) and evaluated yields of pions, emitted copiously both from the
surface of QGP and from the depth of hadronic gas before freeze-out. We
compared their yields with yield of freeze-out pions and show that direct
pions from QGP surface dominate in soft $p_t$ region, while contribution of
pions, emitted from the depth of hadronic gas is much smaller than
contribution of freeze-out pions. Emission of direct pions from the surface
of QGP result in enhancement of the yield of pions with low $p_t$. This
enhancement has been observed long ago in $p_t$ distributions of pions in AA
collisions \cite{pi-low-exp}. A lot of explanations of this excess was
proposed so far. Contributions of several proposed effects appear to be
negligible small, but contributions of the strongest effects (decays of
resonances \cite{pi-low-res}, collective motion \cite{pi-low-coll} and
absence of chemical equilibrium \cite{pi-low-chem}) are sufficient for
explanation of observed enhancement. Therefore, there is strong physical
background for the direct pions from QGP. However, mentioned effects lead to
much smaller contributions into distributions of heavier mesons.

In this letter, in addition to pions, we consider emission of direct
$K,\eta,\rho,\omega$ and $K^*$ mesons in Pb+Pb collisions at 158 $A\cdot GeV$
(SpS) and 3150+3150 $A\cdot GeV$ (LHC) energies and compare them with yields
of corresponding freeze-out mesons. Estimation of situation at RHIC energies
can be obtained by interpolation of our predictions at lower (SpS) and higher
(LHC) energies.

To estimate yields of the direct mesons in heavy ion collision we use the
following model. Hot matter, created in the very beginning of collision,
evolves hydrodynamically. On the background of this evolution the direct
mesons are continuously emitted from the surface of QGP as a result of flying
out of quarks and gluons from the depth of quark-gluon plasma, their
hadronization on the plasma surface and fly out of the direct mesons through
surrounding hadronic gas sometimes with rescattering. If direct meson suffers
rescattering in the hadronic gas, then we assume, that it lost direct
information about plasma and consider it further hydrodynamically. For
freeze-out hadrons we assume thermodynamic and chemical equilibrium at
freeze-out moment.

Probability for quark and gluon being emitted in the depth of QGP to reach
its surface, and for meson to escape from hadronic gas without rescattering
is determined by expression:
$$
P=\exp \left\{ -\int \lambda ^{-1}(\varepsilon ,x)\,dx\right\},
$$
where integration is performed along (straight line) path of the particle in
the hot matter, and $\lambda (\varepsilon ,x)$ - free path length of the
particle, which depends on energy of the particle $\varepsilon$ and local
energy density at the point $x$. We calculate free path lengths of quark and
gluon in QGP and meson in hadronic gas using equation
$$
\lambda _{i}(\varepsilon )=\left[ \frac 1{16\,\pi
^3}\frac T{\varepsilon \ p}\sum_j\int\limits_{(m_1+m_2)^2}^\infty \sqrt{
s^2-2\,s\,(m_1^2+m_2^2)+(m_1^2-m_2^2)^2}\,
\sigma _{ij}(s)\;\ln  \left( \frac{1-\exp (-a_{+})}{1-\exp
(-a_{-})}\right) \,ds\right] ^{-1},
$$
$$
a_{\pm }=\frac{\varepsilon(s-m_1^2-m_2^2) \pm \sqrt{(\varepsilon
^2-m_1^2)( s^2-2\,s\,(m_1^2+m_2^2)+(m_1^2-m_2^2)^2)}
}{2\,m_1^2\,T},\quad
$$
where $\sigma _{ij}(s)$ -- total cross-section of interaction of $i$ and $j$
particles, $m_1$ and $m_2$ -- mass of projectile and target particles
correspondingly, $T$ -- temperature, and sum is taken over all possible
two-particle reactions. Evaluating free path lengths of hadrons in hadronic
gas we take into account only rescatterings on pions. In the case of the free
path length of pions we use experimental cross-sections of $\pi\pi$
scattering (see \cite{We}), while for $K,\,\eta, ...$ we assume
$\sigma=\sigma_{\pi^+\pi^+}\sim 10\, mbarn$ plus contributions from
excitation of resonances. One could expect, that presence of nucleons in the
hadronic gas result in significant reduction of the free path length of the
pion with respect to pure pionic gas (due to excitation of $\Delta$
resonances). However, we find, that for reasonable values of baryonic
chemical potential $\mu\le 200\,MeV$ and temperatures below $\sim250\,MeV$,
contribution of nucleons into free path length of pions is negligible.

Emission rate (the number of particles, emitted from unit volume per unit
time) of quarks and gluons from QGP we find from the condition, that
infinitely thick layer of QGP emits quarks and gluons in accordance with
Stephan-Boltsman formula. So we find:

$$
\varepsilon \frac{d^7R_i}{d^3p\,d^4x}=\frac{d_i}{\lambda
_i(\varepsilon )}\frac \varepsilon {(\exp (\varepsilon /T)\pm 1)}.
$$
where $d_i$ -- degeneracy, $\lambda _i$ -- free path length of the particle,
$T$ -- temperature.

In this letter we are interesting in $p_t$ distributions at midrapidity region,
therefore  we describe evolution of the hot matter by use of Bjorken
hydrodynamics with transverse expansion. To investigate sensitivity of our
predictions to the description of freeze-out process, we consider two extreme
versions of hydrodynamic model with pressure outside freeze-out surface
$P_{outside}=0$ and $P_{outside}=P_{inside}$.

As far as it is not possible to describe consistently hadronization of quarks
and gluons on the plasma surface, we estimate upper and lower limits on the
yields of direct hadrons, by use of two models of hadronization -- `creation'
model, giving the lower limit and `pull in' model, used in the paper
\cite{We} and giving the upper limit. In both models we assume, that a quark
or gluon, flown through the plasma surface, pulls tube (string) of color
field. Creation of the final hadron corresponds to the breaking of the tube
due to discoloring of the moving out quark or gluon. In the first model we
assume, that this discoloring takes place as a result of creation of a
quark-antiquark pair in the strong field of the tube. This assumption is used
in the well known family of event generators JETSET \cite{Jetset}, where all
parameters are chosen to fit the $e^+e^-$ annihilation at $\sqrt{s}=30\,GeV$.
Therefore, in the `creation' model we extract corresponding dependence from
JETSET 7.4. In the second model we assume that discoloring takes place as a
result of `pulling in' of soft quark or gluon with corresponding color from
the pre-surface layer of QGP into the tube. Because of the large number of
soft quarks and gluons in the QGP {\it each} moving out quark and gluon can
transform into some hadron if its energy is larger than mass of this hadron.
If several hadrons can be formed, then we take their relative yields from
JETSET 7.4. Having probabilities of hadronization of quarks and gluons
$R_h^{q,g}(\varepsilon_q)$, evaluated using these two models, we
obtain the following hadronization function:
$$
f_h^{q,g}(\varepsilon_h,\theta_h,\phi_h)=2\, R_h^{q,g}(\varepsilon_q)
\frac{\delta(\phi_h-\phi_q)\,\delta(\cos\theta_h-\cos\theta _q)\,
      \theta(\varepsilon_q-m_h)}
     {\varepsilon_q\sqrt{\varepsilon_q^2-m_h^2}-m_h^2\ln\left
      (\varepsilon_q/m_h+\left .\sqrt{\varepsilon_q^2-m_h^2}
      \right/m_h\right ) }
$$
where $\varepsilon,\theta,\phi$ energy, polar angle and azimuthal angle of
initial quark ($q$) or gluon ($g$) or final hadron ($h$) correspondingly, and
$m_h$ -- mass of the hadron. This fragmentation function is normalized to
describe fragmentation of one quark or gluon to $R_h^{q,g}(\varepsilon_q)$
hadrons with energy in the range $m_h<\varepsilon_h<\varepsilon_q$.

To apply our model for central $Pb+Pb$ collisions at $158\, A\cdot GeV$ we
choose initial conditions (initial energy density $\varepsilon^0_{in}$, time
of thermalization $\tau_{in}$ and initial radius $R_{in}$), QGP-hadronic gas
transition temperature ($T_c$) and freeze-out temperature ($T_{freeze}$) to
reproduce experimental $p_t$ distribution of $\pi^0$ at midrapidity
\cite{SpS-pi0-exp}. Within two versions of the hydrodynamic model, mentioned
above we use:
$
\varepsilon^0_{in}=2.5\,GeV/fm^3, \, \tau_{in}=1\,fm/c, \,
R_{in}=7\,fm,\, T_c=170\, MeV,\,T_{freeze}=130\, MeV
$
for $P_{outside}=P_{inside}$, and
$
\varepsilon^0_{in}=0.35\,GeV/fm^3, \,
\tau_{in}=4.5\,fm/c, \, R_{in}=10.5\,fm, \, T_c=150\, MeV,\,T_{freeze}=140\,
MeV
$
for $P_{outside}=0$. As far as in the case of $P_{outside}=0$ the
significant pressure gradient on the freeze-out surface result in strong
acceleration of freeze-out hadrons, we choose lower initial energy density
and transition temperature and higher freeze-out temperature than in the case
of $P_{outside}=P_{inside}$.

Evaluated $p_t$ distributions of various mesons at midrapidity for the
`creation' (left plot) and `pull in' (right plot) models of hadronization of
quarks and gluons for hydrodynamics with $P_{outside}=P_{inside}$ are shown
on the fig.\ 1. For convenience, distributions of $K,\eta,...$ mesons are
subsequently multiplied by $10^{-2},10^{-4},...$. Solid lines correspond to
direct mesons from QGP surface, and dashed lines -- to freeze-out mesons. We
do not distinguish $\pi^+,\,\pi^0$, and $\pi^-$, $K^+,\,K^-\,K^0$, and $\bar
K^0$, so that we show distributions, averaged over all pions, kaons etc.
Probability of fragmentation of low energy quarks and gluons in the `pull in'
model is higher than in the `creation' model, therefore the number of direct
mesons is higher on the right plot. In the both models of hadronization
contributions of direct mesons increase with increasing of the meson mass,
and in the case of $\eta$ and heavier mesons, direct mesons dominate over
freeze-out ones, for the `pull in' model. Unexpected, that direct mesons,
emitted from the hottest region of the collision, contribute mainly into soft
$p_t$ region. This takes place because during hadronization a quark or gluon
`decays' onto direct hadron and part of the tube which pulls back to the
plasma, so that the temperature of direct hadron significantly decrease with
respect to typical temperatures in the system.

The same distributions, evaluated in the case of $P_{outside}=0$ are
shown on the fig.\ 2. In this case because of the faster fly out of the
hadronic gas the width of the layer of hadronic gas, surrounding QGP, becomes
smaller and it less screens direct hadrons. As a result, the contributions
of direct hadrons become more significant: direct $K,\,\eta$ and heavier
mesons dominate over freeze-out mesons for both models of hadronization.

To investigate dependence of the yields of direct mesons on the energy of
collision, we evaluated these yields in the central Pb+Pb collisions at LHC
energy. We consider case $P_{outside}=P_{inside}$ and use the same values of
$T_c$ and $T_{freeze}$ as for SpS energies and the following initial
conditions:
$ \varepsilon^0_{in}=20\,GeV/fm^3, \, \tau_{in}=1\,fm/c, \,
R_{in}=7\,fm,
$ what corresponds to the multiplicity at midrapidity
$dN^{\pi^0}/dy\sim 2\cdot 10^3$, predicted by cascade models. Distributions
of direct and freeze-out hadrons, evaluated in this case are shown on fig.\
3. In contrast to SpS at LHC energy direct mesons dominate in the soft $p_t$
region for $K$ and heavier mesons for both models of hadronization.

As one can see, contribution of direct mesons enhances the $p_t$
distributions in the soft $p_t$ region. This takes place because during
hadronization of quarks and gluons on the surface of QGP a part of their
energy is pulled back to the QGP and thus direct hadrons have lower
temperature then typical temperature in the system. As far as contribution of
direct hadrons increase with increasing of their mass, we predict increasing
of the low $p_t$ enhancement with increasing of the hadron mass -- in contrast
to all other possible explanations, such as decay of resonances, collective
motion, absence of chemical equilibrium etc. This behavior can be used as
signature of QGP creation.

To conclude, we consider light mesons, emitted directly from the surface of
quark-gluon plasma and escaped from hot matter without rescattering. This
mesons gives unique opportunity to test the pre-surface layer of QGP via
strongly interacting particles. We show, that direct mesons contribute mainly
into soft $p_t$ region ($p_t \le 0.5\,GeV/c$). Using two extreme models of
hadronization of quarks and gluons on the plasma surface and two extreme
models of hydrodynamic expansion, we estimate lower and upper limits on the
yields of direct mesons with different masses. We find, that at SpS energies
in the frame of both models of hadronization direct mesons dominate in the
soft $p_t$ region for $\eta$ and heavier mesons in the case of
$P_{outside}=0$ but only within the `pull in' model of hadronization in the
case of $P_{outside}=P_{inside}$. However, for LHC energy direct $\eta$ and
heavier mesons dominate at soft $p_t$ for all considered models. Contribution
of direct mesons result in appearing of the enhancement of $p_t$ distribution
in the soft region, which can be used as a new signature of the of
quark-gluon plasma creation.

This research was supported in part by grant RFFI 96-15-96548.

\begin{figure}
\caption{Yields of direct (solid lines) and freeze-out (dashed lines) mesons
in Pb+Pb collision at SpS energy within `creation' (left plot) and `pull in'
(right plot) models of hadronization for hydrodynamics with
$P_{outside}=P_{inside}$. Distributions of $K,\eta,...$ are multiplied by
$10^{-2},10^{-4},...$.}
\end{figure}

\begin{figure}
\caption{The same as fig.\ 2, but for hydrodynamics with
$P_{outside}=0$.}
\end{figure}

\begin{figure}
\caption{The same as fig.1, but for Pb+Pb collision at LHC energy.}
\end{figure}

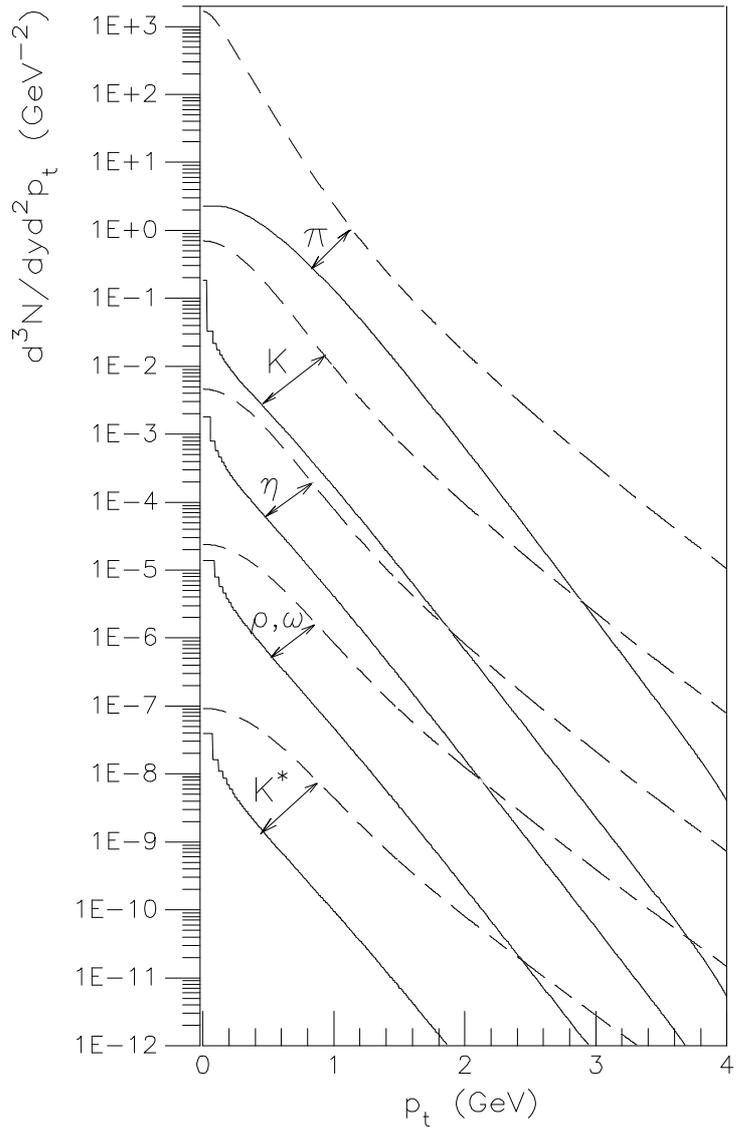
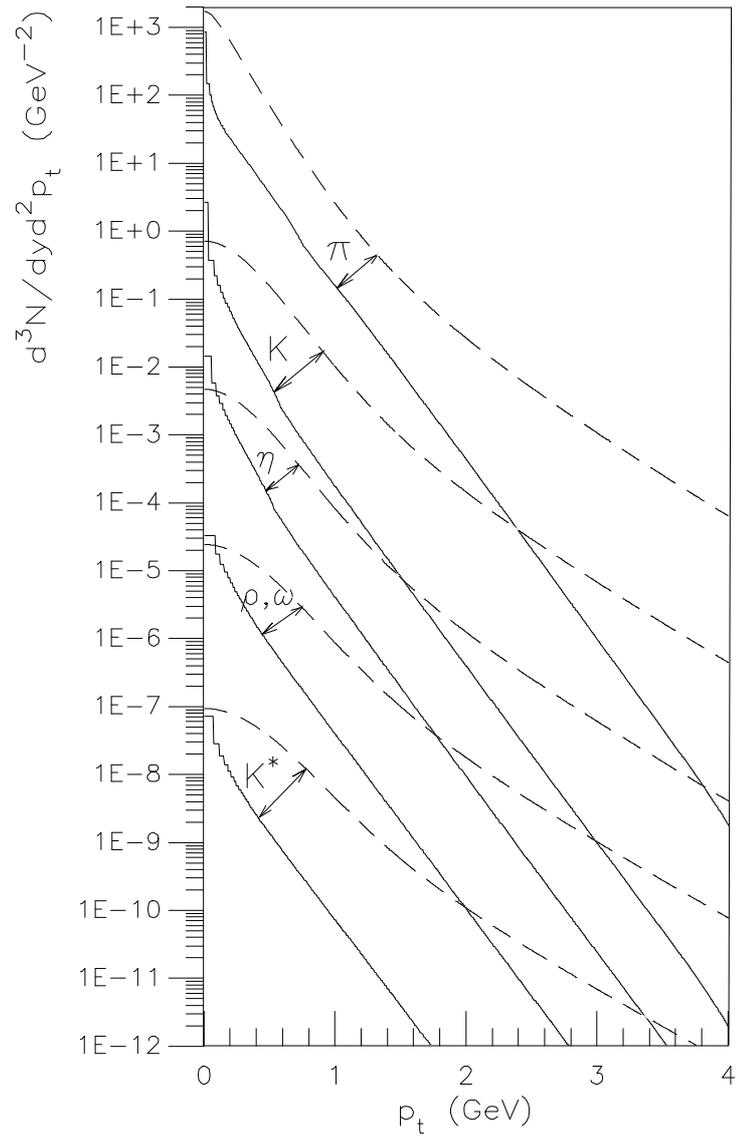
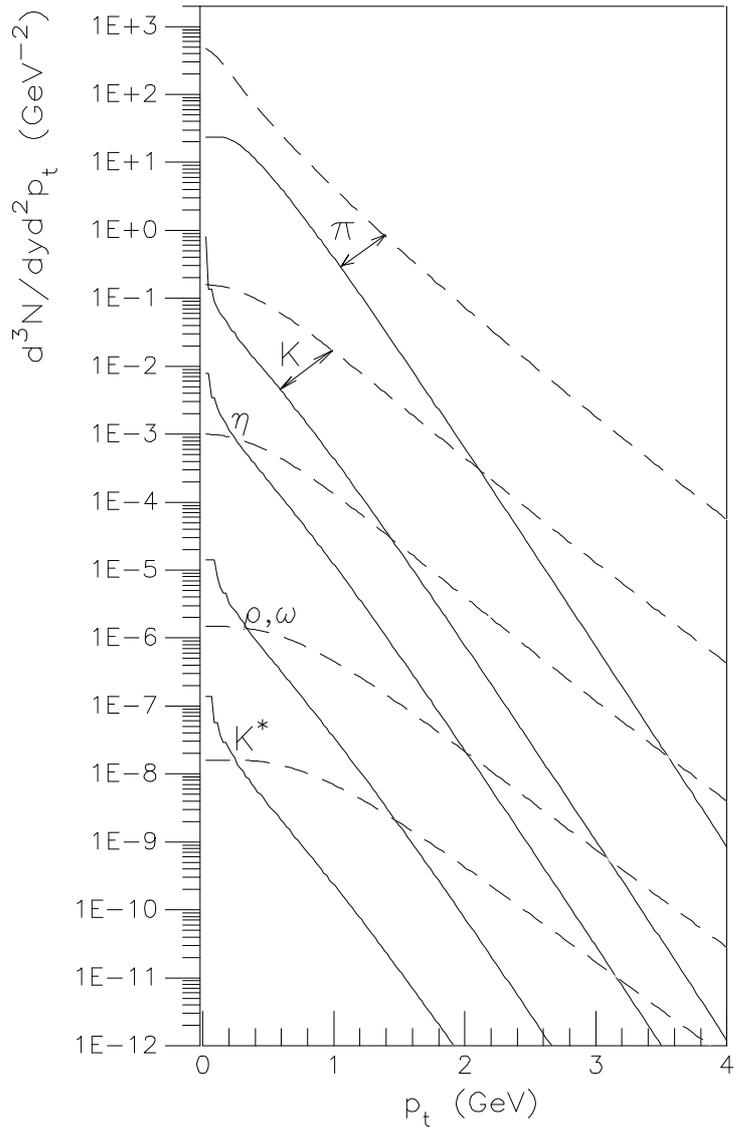
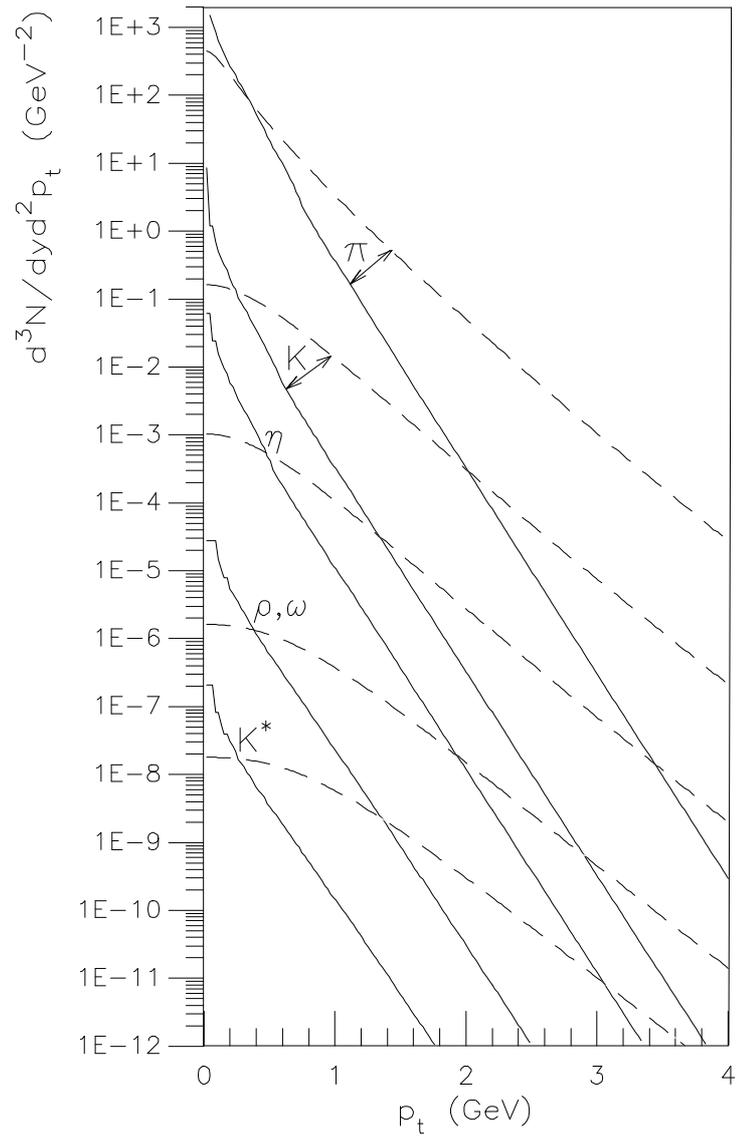
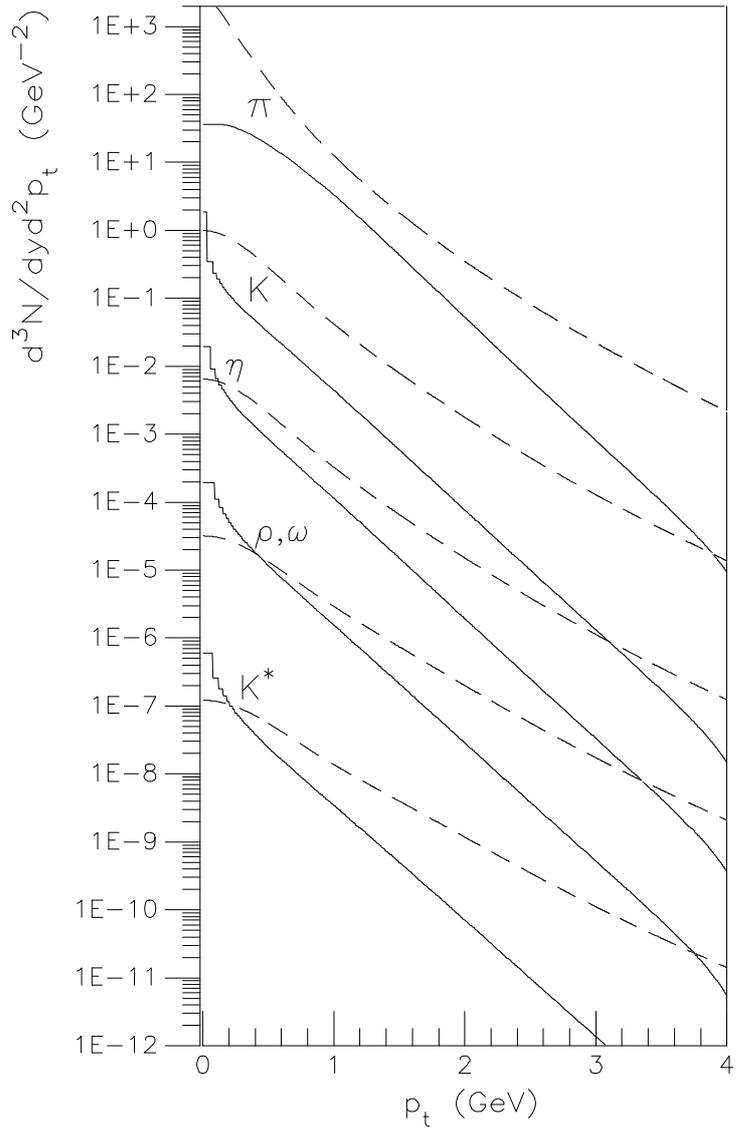
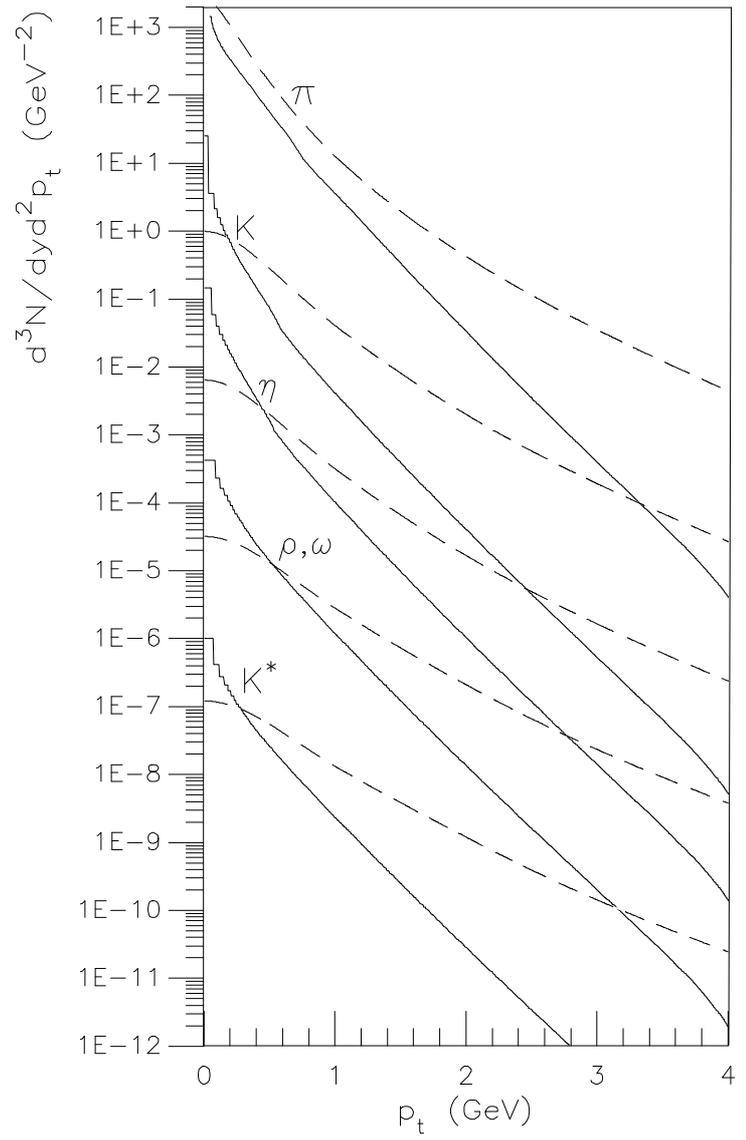
\end{document}